# High-contrast coronagraph for ground-based imaging of Jupiter-like planets[*]

Jiang-Pei Dou[1, 2], De-Qing Ren[1, 2, 3], Yong-Tian Zhu[1, 2]

1 National Astronomical Observatories/Nanjing Institute of Astronomical Optics & Technology, Chinese Academy of Sciences, Nanjing 210042,China; jpdou@niaot.ac.cn
2 Key Laboratory of Astronomical Optics & Technology, Nanjing Institute of Astronomical Optics & Technology, Chinese Academy of Sciences, Nanjing 210042,China
3 Physics & Astronomy Department, California State University Northridge, 18111 Nordhoff Street, Northridge, California 91330-8268

**Abstract** We propose a high-contrast coronagraph for direct imaging of young Jupiter-like planets orbiting nearby bright stars. The coronagraph employs a step-transmission filter in which the intensity is apodized with a finite number of steps of identical transmission in each step. It should be installed on a large ground-based telescope equipped with state-of-the-art adaptive optics systems. In that case, contrast ratios around $10^{-6}$ should be accessible within 0.1 arc seconds of the central star. In recent progress, a coronagraph with circular apodizing filter has been developing, which can be used for a ground-based telescope with central obstruction and spider structure. It is shown that ground-based direct imaging of Jupiter-like planets is promising with current technology.

**Key words:** instrumentation: high angular resolution — methods: laboratory, numerical — techniques: coronagraphy, apodization—planetary systems

## 1 INTRODUCTION

At present, over 300 extra-solar planets have been detected mostly through the radial velocity technique. The detected candidates are either quite massive ($0.16 M_J < M \sin i < 13 M_J$) or very close to their primary stars (0.03-4 AU), whose properties are likely a result of observational bias (Ammons et al. 2006). A numerical simulation suggests that the actual population probably contains many planets including both terrestrial planets similar in size to Earth and giant icy planets in long-period orbits (Ida & Lin 2004). The exact mass of the planets can not be determined through radial velocity technique because of the declination angle. Together with transiting approach, the radius and mass ($1.42 R_J$, $0.69 M_J$) of the planet orbiting HD 209458 are firstly determined (Henry et al. 2000). However, the indirect detection techniques cannot perform spectroscopic measurement and therefore could not determine whether there is a life or not on another planet. Direct detection of an extra-solar planet's emitted or reflected light, would finally provide us complementary of some important physics parameters, such as its chemical composition even the presence of life signals. It can permit us to put constraints on theories of planets formation and migration as well.

It is extremely challenging to direct image an earth-like planet from the ground due to large flux ratio contrast ($10^{-10}$) and its location very close to the primary star (0.1″). For NASA's Terrestrial Planet Finder Coronagraph, a contrast of $10^{-10}$ at an angular distance better than $4\lambda/D$ is required (Brown &Burrows 1990).However, direct detection of the young Jupiter-like planet maybe possible using large ground-based telescopes equipped with state-of-the-art adaptive optics (AO) system (Masciadri et al. 2005; Langlois et al. 2006).Hot Jupiters are supposed to be still self-luminous due to on-going accretion (Marley et al. 2007) and therefore sufficiently bright for direct imaging. In 2008 three Jupiter-size planets around their star HR8799 have been direct detected using 8-m Gemini and 10-m Keck telescopes (Marois et al. 2008). The flux ratio between these planets and their star is around $10^{-5}$, making them possible to be direct imaged from the ground.

[*] Supported by the National Natural Science Foundation of China

These planets were imaged through traditional coronagraph system with AO plus image processing algorithms. Such a system could not reach a very high contrast at present. Introducing a high-contrast imaging coronagraph to the traditional system, more faint extra-solar planets should be directly imaged from ground. In this work, we proposed a high-contrast coronagraph to direct image young Jupiter-size planets. The coronagraph employs a step-transmission filter in which the intensity is apodized with a finite number of steps of identical transmission in each step (Ren & Zhu 2007). The coronagraph is designed to deliver a contrast in the order of $10^{-6} \sim 10^{-7}$ and $10^{-9} \sim 10^{-10}$ as a short and a long term goal, respectively. At present it has demonstrated a stable laboratory experiment result in visible wavelength that a contrast in the order of $10^{-6}$ has been achieved at an angular distance of $4\lambda/D$. The coronagraph should be installed on a large ground-based telescope (6-8 m class) with AO systems to correct the atmosphere turbulence, in which case contrast ratios around $10^{-6}$ should be accessible within 0.1″ of the central star. Three kinds of observation candidates for such a coronagraph system are: a) direct image of the three extra-solar planets that have been detected through traditional imaging coronagraph; b) direct image the planet-hosting stars (PHS) to determine these unconfirmed extra-solar planet candidates; c) direct image some new planets with most favorable contrast of their primary stars (probably PHS).

To direct image the extra-solar planet, a monolithic mirror telescope is preferred to eliminate the diffraction of each segment mirror. However, ground-based telescopes are not an off-axis design. The central obstruction and spider structure will introduce a serious diffraction, which will greatly degrade the performance of existing coronagraph systems. To overcome such a problem, we have been developing a coronagraph based on the circular-step-transmission filter that is suitable for a ground-based telescope with central obstruction and spider structure. Such a coronagraph can effectively suppress the diffraction lights along the four diagonal directions according to the theoretical simulation. The target contrast for the newly designed coronagraph is set to be $10^{-6.5}$ at an angular distance around $5\lambda/D$. With such a contrast, the new coronagraph system will possess a potentiality to direct image Jupiter-size planets based on current ground-based telescope such as Subaru or Gemini.

The outline of the paper is as following. In Section 2, the principle of operation is presented. In Section 3, we propose the development of the step-transmission coronagraph. In Section 4, the potential extra-solar planet candidates to observe are discussed. The conclusions and future developments are given in Sections 5.

## 2 PRINCIPLE OF OPERATION

The light of star combined with that of planet candidates from the sky will enter the following basic instruments:

Telescope->AO system->Coronagraph system->Detector

To direct image the extra-solar planet that is very close to its star, a 6~8m class telescope is preferred for such a ground-based observation which can gain a high spatial resolution. On the other hand, the planet delectability is also affected by the halo of scattered light which mainly comes from atmosphere turbulence. Detection of planet that is roughly 15 mag fainter than its host, requires a halo $10^4$ times fainter than the central image peak. Introducing a high precision AO system can greatly reduce the diffraction halo to a great amount (Langlois et al. 2006). After AO correction lights will enter the coronagraph system that is composed of collimated and imaged mirrors, high-quality step-transmission filter and cross-shaped mask. At a given wavelength, the star diffraction lights can be effectively suppressed to the point spread function (PSF) image center. As a result, most of the energy will be distributed in the central part of the PSF image, making the planet's faint light locating several arc seconds to its star detectable.

The main advantage of the step-transmission filter based coronagraph lies in the fact that it can reach a much higher dynamic range than traditional coronagraphs. The intensity of its step-transmission filter is apodized with a finite number of steps of identical transmission in each step. Such a design simplifies the transmission pupil manufacture and high precision of transmission could be achieved by adopting the "measurement before coating" procedure (Ren & Zhu 2007). Another advantage of the step-transmission filter based coronagraph is its high throughput over 40%, which will greatly reduce the exposure time for a practical observation.

## 3 DEVELOPMENT OF THE STEP-TRANSMISSION COROANGRAPH

### 3.1. General Requirements for a Coronagraph

For ground-based direct imaging of young Jupiter-size planets, a coronagraph must provide high-contrast imaging with a contrast from $10^{-4}$ to $10^{-6}$, applicable both in visible and near infrared (NIR) wavelengths. Meanwhile, high spatial resolution is another critical requirement for the coronagraph to direct image planet very close to its primary star. For our coronagraph, the inner working angle (IWA) is set to be around 4 $\lambda/D$. Correspondingly, for an 8-m class telescope, the detectable planet distance to its star will be 0.05″ in visible (0.5μm) and 0.2″ in the NIR (2μm) wavelength, respectively.

### 3.2 Principle of the Coronagraph for Extra-solar Planets Imaging

The coronagraph is still based on the principle of a step-transmission filter as proposed by Ren and Zhu two years ago (Ren& Zhu 2007) but some slight changes have been made on the new filter development. Here we briefly review the principle of the coronagraph.

The electric field of the electromagnetic wave at the pupil plane for the coronagraph system can be expressed:

$$E_{pupil}(u,v) = A(u,v), \qquad (1)$$

where $A(u, v)$ is the entrance pupil function of the coronagraph system.

The electric field of the starlight on the focal plane is the Fourier transform of the electric field on the pupil plane. The associated intensity of the PSF image is the square of the complex modulus of the electric field on the focal plane and is given as:

$$I(x,y) = \left| \vec{F}[E_{pupil}(u,v)] \right|^2, \qquad (2)$$

where $\vec{F}$ represents the Fourier transform of the associated function.

The intensity distribution of the PSF image can be changed through choosing different pupil functions. In this work, the pupil of the coronagraph is apodized with a finite number of steps of identical transmission in each step. Through optimizing the transmission for each step of the pupil, the most energy of the focal plane image will be distributed in the PSF center accordingly.

We have identified several contrast-limiting errors in recent laboratory experiments (Dou et al. 2009). Among them is the "none-straight" shape of the overlap between two neighboring steps, which is a limitation of the existing coating procedure. As a result, several bright spots randomly occurred in the PSF image. Correspondently the contrast in these regions can not reach its theoretical result. To overcome the problem, a thin straight stripe with no light pass is introduced between each two neighboring steps, which will restrict the diffraction along one direction (for instance, diffraction can be only along the vertical direction with a horizontal apodizaiton). Such design will not affect the theoretical contrast of the coronagraph. A contrast in the order of $10^{-6}$ should still be achieved at an angular distance of 3~4$\lambda$/D. However, the throughput will reduce a little bit from 41.5% to 36.1% due to existing stripes. In this work, we increase the step number from 13 to 15 to gain a larger outer working angle (OWA), which is roughly equal to $\lambda$/D times the step number of the filter. Figure 1 shows the transmission amplitude pattern of the newly designed filter and photograph of the actual filter.

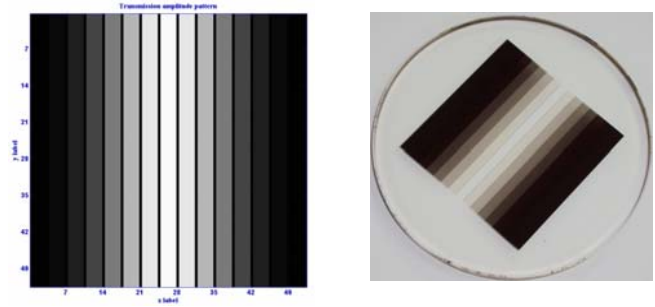

**Fig.1** Left: Transmission amplitude pattern of the newly designed filter, a 50 um straight stripe is induced between each two neighboring steps; Right: photograph of the actual filter.

**3.3 Recent Laboratory Experiment Results**

In this sub-section we present the latest laboratory experiment results of the coronagraph in visible wavelength. The coronagraph employs a new 15 step-transmission filter according to the design discussed above (see Figure 1). The configuration of the experiment system can be found in recent paper (Dou et al., 2009). Figure 2 shows the PSF images under different exposure times of 0.09, 2.7, 36 and 360s, respectively. It is shown that these randomly distributed speckles have been effectively removed and the diffraction is restricted along the vertical direction due to the straight stripes. As a result, a high contrast of $10^{-6}$ has been achieved at an angular distance of 4$\lambda$/D along the diagonal direction. The tested contrast plot along the diagonal direction is shown in Figure 3.

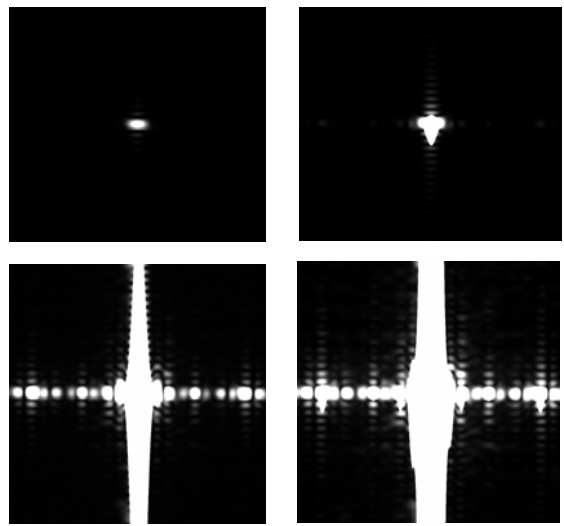

**Fig.2** PSF images under different exposure times.

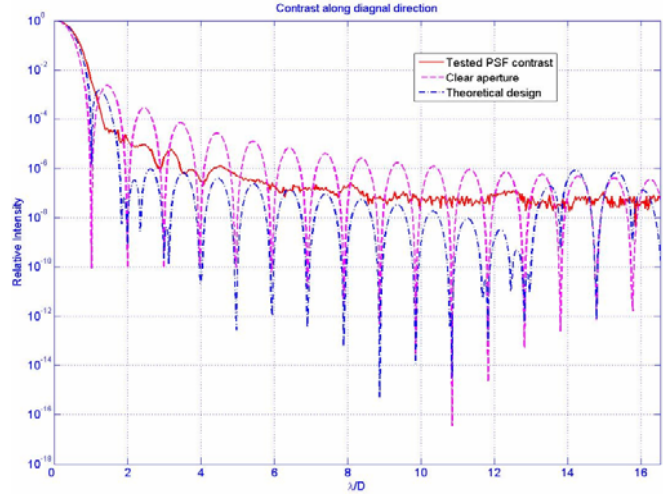

**Fig.3** Tested contrast plot along diagonal direction for clear aperture (dashed purple line), theoretical filter apodization (blue dotted line) and practical modulation (red real line).

**3.4 Coronagraph Developing for Ground-based Observation**

At present, ground-based telescopes are not an off-axis design. The central obstruction and spider structure will introduce a serious diffraction, which will greatly degrade the actual performance of existing coronagraph systems. To overcome the problem, we have been developing a coronagraph that can be used for a ground-based telescope with central obstruction and spider structure. In this sub-section we briefly present its theoretical design and performance.

Here we set the central obstruction of telescope to be 12.5%. And the width of the spider arms is set to be 1.6% of the primary mirror diameter (For instance, 13 cm for a telescope with an 8-m primary mirror, which is thick enough). In this work, the coronagraph employs a step-transmission filter with finite circular steps. In each circular step, the transmission value is the same. The employment of step-transmission filter significantly simplifies the manufacturing of the transmission pupil, making high-precision transmission achievable for the filter.

Using step-transmission filter to achieve high-contrast imaging, our goal is to find the optimum transmission value $T_n$ for each circular step, where n is step number. The optimization problem involves minimizing the energy in a target region R and maximizing total throughput as well. It can be formulated as a 2-D constrained nonlinear minimization problem:

minimize $$\sum_{(x,y)\in R}|C(x, y - C_t)|,$$

subject to $0 \leq T_n \leq 1$, (3)

where $C$ is the approximating contrast of the system and $C_t$ is the target contrast.

Here we normalize the intensity of the PSF image by dividing the intensity maxima, and then we gain the contrast:

$$C(x, y) = \frac{I(x, y)}{\max[I(x, y)]}.$$ (4)

The target contrast for the newly designed coronagraph is set to be $10^{-6.5}$ at $5\lambda/D$. The solution obtained for

IWA= 5λ/D and OWA=135λ/D is shown in Figure 4. The throughput is 32.5%. It is shown that the extra diffraction caused by central obstruction and spider structure has been effectively suppressed along the four diagonal directions. To demonstrate its theoretical performance, we show in Figure 5 the PSF contrast plot from 5λ/D to 50λ/D.

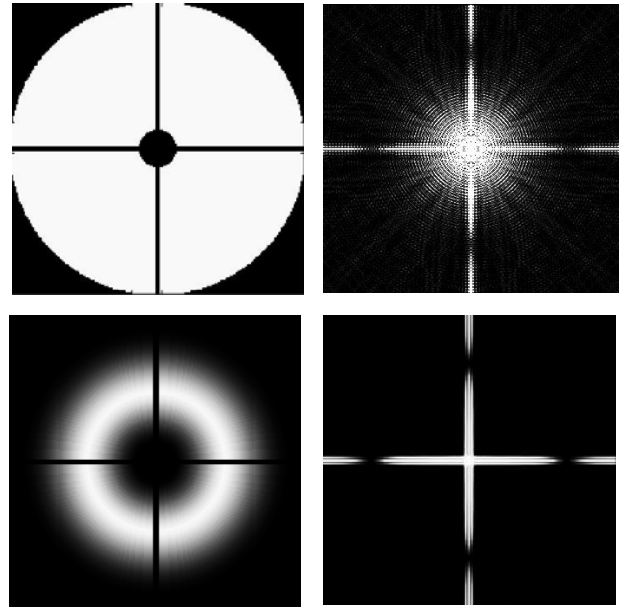

**Fig. 4** Upper: The pupil shape of a telescope with central obstruction and spider structure (left); the associated PSF image with a serious diffraction due to telescope structure (right). Bottom: Amplitute pattern for the circular-step-transmission filter (left); associated PSF with four dark region achieved along diagonal directions (right).

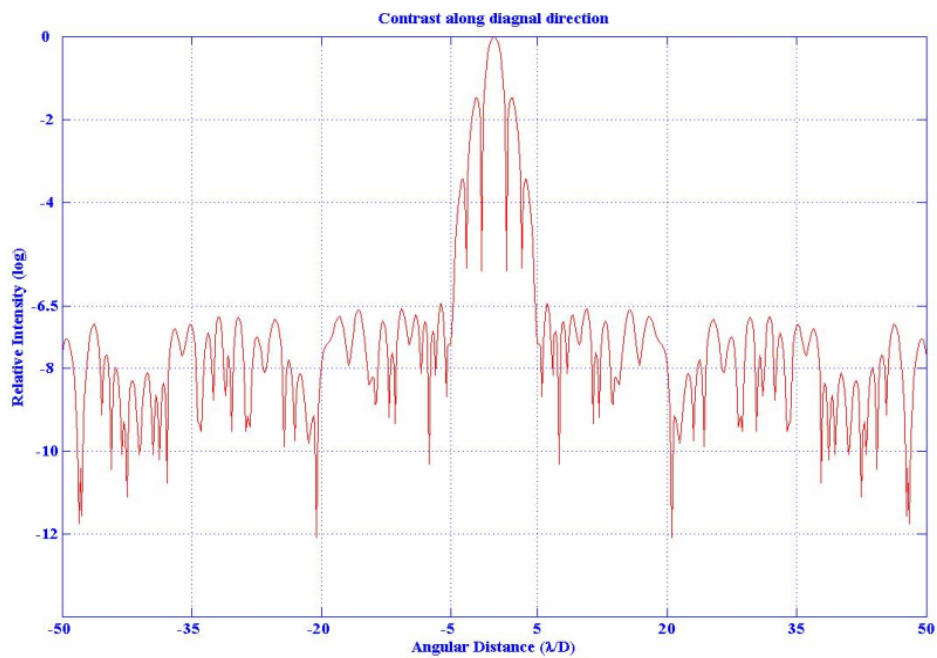

**Fig. 5** PSF contrast plot: a contrast in the order of $10^{-6.5}$ can be achieved in the four diagonal regions.

# 4 POTENTIAL EXTRA-SOLAR PLANET CANDIDATES

## 4.1 Candidates

Our coronagraph should be installed on a large ground-based telescope (6-8 m class) with AO systems to correct the atmosphere turbulence. In that case contrast ratios around $10^{-6}$ should be accessible within 0.1 arc seconds of the central star. To testify the feasibility of such a coronagraph for ground-based observation, we plan to use it firstly to direct image the three detected planets (Marois, et al. 2008). Since the location of these planets has been already confirmed, it is much easier for us to know the actual performance of the coronagraph quickly. Table 1 shows the basic properties of the three planets around its primary star HR8799. Using the coronagraph higher contrast should be achieved than traditional coronagraph systems and therefore these planets are supposed to be seen much clearly.

Table 1 Basic properties of the HR8799 system

| Star | Distance from Earth | $T_{eff}$ [K] | | |
|---|---|---|---|---|
| HR8799 | 38~40pc | 7505-7305 | | |
| Planets | Distance (Angular) | $T_{eff}$ [K] | Contrast | Wavelength |
| HR8799 b | 24AU(0.63") | 900-800 | $10^{-7}$~$10^{-5}$ | J,H,K (1.25,1.65,2.2μm) |
| HR8799 c | 38AU(1") | 1100-1000 | $10^{-6}$~$10^{-4.5}$ | |
| HR8799 d | 68AU(1.8") | 1100-1000 | $10^{-6}$~$10^{-4.5}$ | |

As a following application our coronagraph could be used to image the extra-solar companion that has not been confirmed as planets. In Lagrange's recent paper a probable giant planet has been imaged in the β Pictoris disk by using AO plus image processing algorithms. Such companion could be the first extra-solar planet ever imaged so close to its primary star. (Lagrange et al. 2009). Inducing the coronagraph to existing system would provide higher contrast ($10^{-5}$~$10^{-6}$) and spatial resolution (within 0.2"), which will help to determine the planet candidate. On the other hand, some planets' mass determination is obtained through theoretical models rather than from observation (Neuhauser et al. 2005). Using such a coronagraph, direct imaging of these planets becomes promising and their actual mass and other important physics parameters could be precisely measured finally. Table 2 shows some physical properties of the planets or planet candidates (Ducourant A&A 2008; Lagrange et al. 2009).

Table 2 Basic parameters of planets or planet candidates up to decide

| Planets | Distance (Angular) | $T_{eff}$ [K] | Age [Myr] | Mass[$M_{jup}$] |
|---|---|---|---|---|
| GQ Lup b | 103AU(0.7") | 2520-1600 | <=2 | 1-50 |
| 2M1207 b | 46AU(0.8") | 2000-1100 | 8 | 3-10 |
| AB Pic b | 275AU(6") | 2400-1600 | 30-40 | 11-70 |
| β Pic b | 8 AU (0.4") | 1600-1400 | 8-20 | 6-12 |

Another probable application for our coronagraph is to direct image the nearby region of those favorable stars. As Mark Marley suggested that the most favorable contrast would arise from a young, massive Jupiter orbiting an M star. We may use our coronagraph to direct image regions around some of the young M stars. In that case, we hope to discover some new planets that have never been detected through existing techniques, which will enrich the population of the extra-solar planets.

**4.2 Observation Wavelength**

To demonstrate the optimum wavelength for observation, we calculate the flux ratio between the three ever-detected planets and their primary star HR8799. These planets are very young and still self luminous hence we use the black-body radiation theory to make the simulation. Figure 6 shows simulation results of ratio contrast in different wavelengths, which are consistent with the observation results. For HR8799 c and d, the contrast ratio is about $10^{-4.5}$ in K band and $10^{-6}$ in J band, respectively. It is obvious that the contrast difference is lower in IR than in visible wavelength, which makes long wavelength observation favorable. On the other hand, the spatial resolution will reduce when increasing the wavelength since the IWA of the coronagraph is proportional to $\lambda/D$. For an 8-m class telescope, the detectable planet distance to its star will be 0.05″ in visible and increase to 0.2″ in K band. Although direct imaging of planets in Mid-IR is much easier than in NIR from the ratio contrast view, we will use the coronagraph to direct image the extra-solar planet close to its parent star in NIR wavelength.

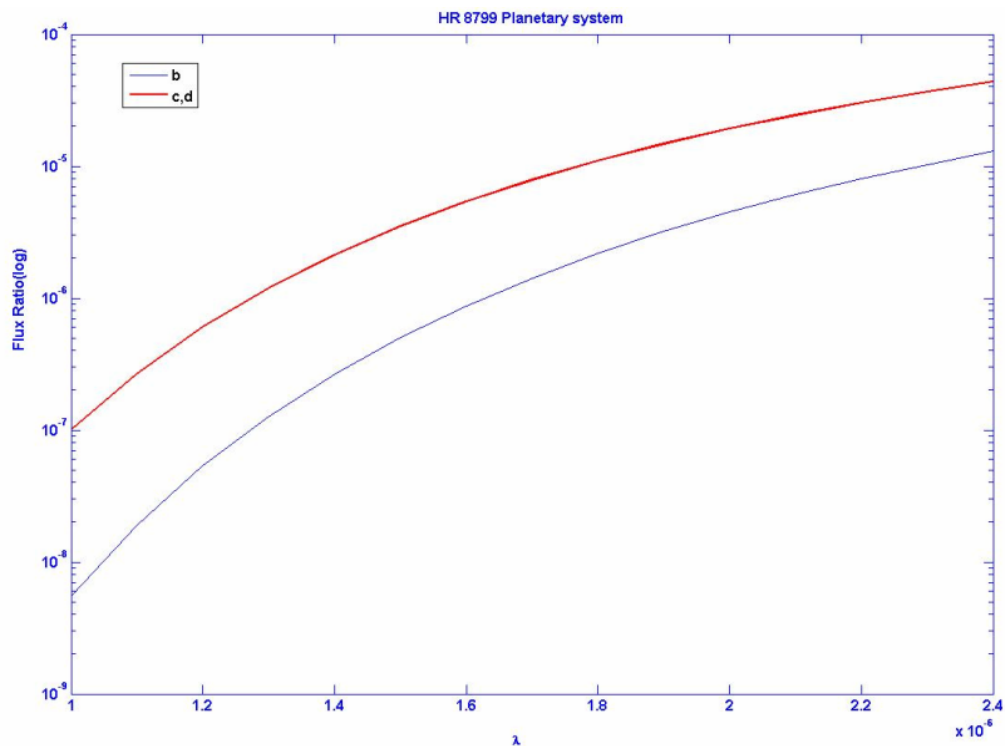

**Fig.6** Simulation flux ratio for the HR 8799 system in different wavelengths.

**5 CONCLUSIONS AND FUTURE DEVELOPMENTS**

In this work, we show that ground-based direct imaging of young Jupiter-size planets is promising with current technology. Now a step-transmission filter based coronagraph has been developed in the institute. It has reached a stable contrast in the order of $10^{-6}$ at $4\lambda/D$ in the visible wavelength, which is one of the best laboratory results compared with other research groups in the world. Using such a coronagraph direct imaging of Jupiter-like planet around its nearby stars in the NIR wavelength becomes possible. In recent progress, we have been developing a coronagraph which is suitable for the ground-based telescope with central obstruction and spider structure. Numerical simulation shows that such coronagraph can reach a mid-high contrast in the four diagonal regions. In this paper we mainly discuss the coronagraph optimized for ground-based telescopes with monolithic mirror. The coronagraph that can be used for segmented mirror telescopes will be discussed in future works.

At present, the actual performance of the coronagraph is limited by wave-front error caused by imperfection manufacture of the optical components for an optics system, which induces speckle noise (Ren & Wang 2006). The wave-front error is a main contrast limiting factor and therefore the current coronagraph can not reach a contrast better than $10^{-7}$ in laboratory. As a follow-up effort, we will introduce the simultaneous differential imaging technique to remove speckles in order to gain an extra-contrast of $10^{-2}$. In a long term run, a deformable mirror that has been manufactured will be used to induce specific phase to the coronagraph system (Dou et al. 2009), which should deliver a better performance with a contrast of $10^{-10}$. Such a coronagraph will be installed on a space-based telescope and be finally used for direct imaging of Earth-like planets. Later results will be discussed in our future publications.

**Acknowledgements** This work is supported by the National Natural Science Foundation of China (NSFC) under Grant No. 10873024. We would like to thank the Mirror Laboratory of the institute for assistance with the substrate manufacture. We will greatly thank Changchun Institute of Optics, Fine Mechanics and Physics for coating the filter. We are grateful to thank Dr. Mark Marley for his kindly suggestions on what type of star to observe.

**References**

Ammons, S. M., Robinson, S. E., Strader, J., et al. 2006, ApJ, 638, 1004
Brown, R. A., & Burrows, C. J. 1990, ICARUS, 87,484.
Dou, J. P., Ren, D. Q., Zhu, Y. T., et al. 2009, Sci. China Ser G, 52, 1284
Dou, J. P., Ren, D. Q., Zhu, Y. T., et al. 2009, Proc. SPIE, 7440, 744019
Ducourant, C., Teixeira, R., Chauvin, G., et al. 2008, A&A, 477, L1
Henry, G. W., Marcy, G.W., Butler, R.P. et al. 2000, ApJ, 529, L41
Ida, S., & Lin, D. N. C. 2004, ApJ, 616, 567
Lagrange, A.M., Gratadour, D., Chauvin, G., et al. 2009, A&A, 493, L21
Langlois, M., Burrows, A., & Hinz, P. 2006, A &A, 445, 1143
Marley, M. S., Fortney, J. J., Hubickyj, O., et al. ApJ, 2007, 655,541
Marois, C., Macintosh, B., Barman, T., et al. 2008 Science, 322, 1348
Masciadri, E., Geissler, K., Kellner, T., et al. 2005, IAU, 200,501
Neuhauser, R., Broeg, C., Mugrauer, M., et al. 2005, Proc. IAU, 200, 41
Ren, D.Q., & Zhu, Y.T. 2007, PASP, 119, 1063
Ren, D. Q., & Wang, H. M. 2006, ApJ, 640, 530